\documentclass[aps,prb,twocolumn,showpacs,amsmath,amssymb,floatfix]{revtex4}
\usepackage{epsfig}
\usepackage{bm}
\def\be{\begin{equation}}
\def\ee{\end{equation}}
\def\la{\langle}
\def\ra{\rangle}

\begin{document}
\title{Mesoscopic threshold detectors: Telegraphing the size of a fluctuation}
\author{Andrew N. Jordan and Eugene V. Sukhorukov}
\affiliation{D\'epartement de Physique Th\'eorique, Universit\'e de Gen\`eve,
        CH-1211 Gen\`eve 4, Switzerland}
\date{April 1, 2005}

\begin{abstract}
We propose a two-terminal method to measure shot noise in mesoscopic
systems based on an instability in the current-voltage characteristic
of an on-chip detector.  The microscopic noise drives the instability,
which leads to random switching of the current between two values, the
telegraph process.  In the Gaussian regime, the shot noise power
driving the instability may be extracted from the I-V curve, with the
noise power as a fitting parameter.  In the threshold regime, the
extreme value statistics of the mesoscopic conductor can be extracted
from the switching rates, which reorganize the complete information
about the current statistics in an indirect way, ``telegraphing'' the
size of a fluctuation.  We propose the use of a quantum double dot as
a mesoscopic threshold detector.
\end{abstract}
\pacs{73.23.-b,05.40.-a,74.40.+k,72.70.+m}

\maketitle

\section{Introduction}
Shot noise in mesoscopic conductors,\cite{BB,BS} has attracted great
interest, both theoretically and experimentally.  While shot noise
measurements are an important tool in experimental labs, the
measurement of non-Gaussian noise presents an experimental
challenge.\cite{reulet,R} Noise characteristics beyond the size of the
typical fluctuation, known as full counting statistics
(FCS),\cite{FCS1,FCS2,FCS3} are interesting because they bring
additional information about the transport properties of the measured
conductor.  In particular, the extreme value statistics (EVS) can
have qualitatively different behavior than typical
fluctuations,\cite{nazarovJJ,us0,SB} and thus gives rise to new physical
effects. How these rare fluctuations can be measured is one
outstanding question that this paper is concerned with, and has only
recently received attention.\cite{nazarovJJ,PekolaJJ}

The standard measurement method runs the current through a series of
cables, filters and amplifiers before the noise is detected.  While
this works well for the noise power, and can be extended to the third
cumulant with great effort,\cite{reulet,R} it is very hard to
experimentally measure rare current fluctuations.  A breakthrough in
measurement technology came with on-chip detectors, which use
superconducting devices, or quantum dots
for a variety of functions, such as fast qubit 
read-out\cite{onchip,rmp} or
high-frequency quantum noise measurement.\cite{noiseDD,schoelkopf}

In addition to the many advantages of going on-chip, a further
possibility advanced in this paper is the use of two-terminal, rather
than four-terminal measurements for low frequency noise.
Four-terminal measurements are intrinsically limited by the small coupling
constant between the measurement circuit and the conductor, as well as by the
fact that the low-frequency noise can evade the measurement device by leakage
through the bias line.\cite{pekolaexp} In contrast, a two-terminal
noise detector is strongly coupled, and detection is fundamentally a
non-perturbative process that serves as a preamplifier of the low
frequency microscopic noise.

\begin{figure}[t]
\begin{center}
\leavevmode
\psfig{file=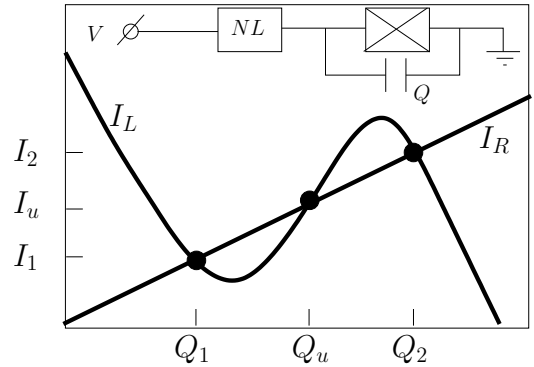,width=7cm}
\caption{The currents of the circuit elements are plotted as a
function of the charge on the capacitor $Q$.  The left element (NL)
has a region of negative differential resistance, allowing bistability:
The average current in the circuit is conserved at two stable points,
$Q_{1,2}$, and one unstable point, $Q_u$.  Inset: Noise measurement
circuit.  The mesoscopic conductor with parallel capacitor
are connected in series with the nonlinear element.}
\label{IVcircuit}
\end{center}
\vspace{-5mm}
\end{figure}
In order to exploit the above advantages, we propose circuits with an
instability as detectors of low frequency noise, as well as
FCS.  The considered on-chip circuit consists of a mesoscopic
conductor with a parallel mesoscopic capacitor, connected in series 
with the nonlinear element (see Fig.~1).\cite{us0} The
nonlinear element has a region of negative differential resistance,
which allows bistability.  The mesoscopic conductor loads the
instability, so that there are two stable charge points on the
capacitor, corresponding to two different currents through the
circuit.  In this bistable range, the shot noise occasionally causes
the circuit to transit from one stable state to the other, producing a
random telegraph signal.\cite{machlup} The rate of transition is
exponentially sensitive to the size of the fluctuation,\cite{us0} and
thus serves as a threshold detector for the rare current fluctuations.  
Although the threshold rates are not a direct measurement of
the FCS, they reorganize the complete information about the noise
statistics in an indirect way, ``telegraphing'' the size of a
fluctuation.  Therefore, bistable systems are a promising candidate for
low-frequency noise detectors, and can confirm or falsify a given
prediction for FCS.

The paper is organized as follows.  In Sec.~\ref{BS} we first review
the statistical properties of the general telegraph process, as well
as the instanton dynamics for noise driven circuits, and the results
for the bistable switching rates.  This sets the stage for the
application of this physical process.  The implications of the
Gaussian noise limit are considered in Sec.~\ref{gaussian}, and we
find that the I-V curve may be used to extract the noise power as a
fitting parameter.  The results for the second and third cumulant are
also given, and the role of the asymmetry in the rates is discussed.
In Sec.~\ref{CE}, we examine several effects that arise when the
bistable circuit is combined with an external circuit.  The threshold
detector of EVS is introduced in
Sec.~\ref{thresh}.  We give results for the threshold rates of several
processes, and discuss stabilization effects that arise when 
the tails of the distribution have a cut-off. The quantum double dot is
proposed as an implementation of the threshold detector in
Sec.~\ref{DD}.  We discuss the mesoscopic circuit, needed conditions and
constraints, as well as feasibility.  Sec.~\ref{conc} contains our
conclusions.

\section{Set-up and bistability results}
\label{BS}
We first review the essential results on the transport statistics of
bistable systems.\cite{us0} Consider the circuit shown in the inset of
Fig.~1, biased with voltage $V$. The average current through both
circuit elements is plotted in Fig.~1 versus the charge on the
parallel capacitor $C$ (we choose to speak about the charge on the
capacitor, rather than the voltage across the right element $Q/C$.)
The nonlinear element on the left has a range of negative differential
resistance, which leads to the possibility of three charge/current
points, $(Q_1, I_1), (Q_2, I_2), (Q_u, I_u)$, where the I-Q curves
intersect, so the average current is conserved in the circuit.
The central intersection at $Q_u$ is unstable to small charge
perturbations, while the outer two intersections $Q_{1,2}$ are stable.
The microscopic non-equilibrium noise is correlated on a short time
scale $\tau_0\sim\hbar/eV$, and drives the collective system on the
longer RC-time of the circuit, $\tau_C \gg \tau_0$.  Occasionally, the
microscopic noise causes the system to transit between stable states.
As a result, the measured current switches back and forth between
$I_1$ and $I_2$, with rates $\Gamma_{1,2}$.  These rates contain
valuable information about the statistical nature of the driving noise
that will be examined later.  On a long time scale, the system relaxes
with the rate $\Gamma_S=\Gamma_1+\Gamma_2$ to the stationary state.
This stationary state has constant probabilities to occupy one of the
two stable points, \be P_1=\Gamma_2/\Gamma_S,\quad
P_2=\Gamma_1/\Gamma_S.
\label{state}
\ee 
Therefore, the average current is
\be
\langle I\rangle = \sum_{n=1,2}I_n P_n.
\label{c1}
\ee
The randomness of the duration in either of the stable states leads to the
fluctuation of the transmitted charge
 \be {\cal Q}(t) = \int_0^t dt' I(t').
\label{q}
\ee This random variable has a probability distribution $P({\cal
Q}(t))$, which may be specified by its moments.  In the stationary
limit, it is more convenient to consider the cumulants
(irreducible correlators) because they are linear in time, $\la\la
{\cal Q}^n \ra\ra \equiv t \la\la I^n \ra\ra$, and may be used to
define time-independent current cumulants, $\la\la I^n \ra\ra$.  The
second cumulant and third cumulant of the switching current described
above are given respectively by\cite{us0}
\begin{eqnarray}
\langle\langle I^2\rangle\rangle &=&
 \sum_{n=1,2}F_nP_n + 2 (\Delta
 I)^2\Gamma_1\Gamma_2/\Gamma_S^3,\label{c2} \\
 \langle\langle I^3\rangle\rangle &=&  \sum_{n=1,2}L_nP_n
+ 6(\Delta I)^3\Gamma_1\Gamma_2 \Delta\Gamma/\Gamma_S^5, \label{c3}
\end{eqnarray}
where $\Delta I=I_2-I_1$, while $F_n$ and $L_n$ are the noise power and third
cumulant of the stable points, which describe the small fluctuations
around $I_{1,2}$.  The first term in Eq.~(\ref{c2}) is the weighted
noise power of the stationary states, and the second term is the well-known
result for zero-frequency telegraph noise. \cite{machlup}
The telegraph contribution dominates the bare contribution because it
scales as $\Gamma_S^{-1}$ in the second cumulant, and as
$\Gamma_S^{-2}$ in the third.\cite{us0}

An important feature of bistable systems is that the I-V curve makes a
rapid transition from $I_1$ to $I_2$ as a function of the bias voltage
(see Fig.~2 and the discussion below), while the current cumulants
show a peak structure. During this transition, the specific values of
the currents $I_{1,2}$ may be considered constant.  If one has access
to the first three cumulants of an unknown process, one may utilize
these cumulants, Eqs.~(\ref{c1},\ref{c2},\ref{c3}), to diagnose
whether bistability exists or not.  The dominant telegraph
contribution to Eqs.~(\ref{c1},\ref{c2}) may be used to eliminate the
rates, and substitution into Eq.~(\ref{c3}) then yields the third
cumulant of the telegraph process in terms of the first
two:\cite{note1} 
\be \langle\langle I^3\rangle\rangle_{tel} = 3
\langle\langle I^2\rangle\rangle_{tel}^2 \frac{
(I_1+I_2)/2-I}{(I_2-I)(I-I_1)}.
\label{third}
\ee 
This equation may serve as a valuable test of experimental data
because there is no fitting parameter.  The above 
procedure was used by Flindt, Novotny, and Jauho to demonstrate
bistability in numerical studies of the nanomechanical
shuttle.\cite{shuttle}

The above results (\ref{c1},\ref{c2}-\ref{third}) are general and apply to
any telegraph process, independent of its microscopic origin.
We now turn to the bistable circuit driven by current noise.
The microscopic current
fluctuations $\tilde I_{L,R}$ of the two circuit elements may be
described with generating functions of the current cumulants
(that are Markovian after the correlation time $\tau_0$),
\be H_{\alpha}(\lambda_{\alpha})=\sum_n
(\lambda_{\alpha}^n/n!)  \la\la{\tilde I}^n_\alpha\ra\ra,
\quad\alpha=L,R,
\label{generators}
\ee where the cumulants $\la\la{\tilde I}^n_\alpha\ra\ra$ are
functions of the charge on the capacitor $Q$ (we set the electron 
charge e=1 throughout the paper).
The fact that the correlation time $\tau_0$ is much smaller than the
RC-time $\tau_C$ of the circuit, means the slow dynamics is
classical.\cite{us1} The circuit dynamics may now be described with
the stochastic path integral formalism.\cite{us1,us2} This formalism
is quite general and has been applied to a wide variety of stochastic
problems in mesoscopic physics.\cite{SB,us0,spi1,spi2,spi3,spi4}
In addition to the separation of time scales, we require
that the instability is well developed, so that
the stochastic bistable switching rates are given by\cite{us0}
\be \Gamma_{1,2} = \omega_{1,2} \exp(-A_{1,2}),
\label{delta}
\ee
where the action
\be A_{1,2}= \int^{Q_u}_{Q_{1,2}} d Q\, \lambda_{\rm
in}(Q),
\label{act1}
\ee
must be larger than one.
The attempt frequency $\omega_{1,2}$ is subdominant and will
be neglected.
The function $\lambda_{\rm in}(Q)$ (which we
refer to as the instanton line) is implicitly
defined by the nontrivial solution of the algebraic equation\cite{us0}
\be
H(Q,\lambda) = H_L(Q,\lambda)+ H_R(Q,-\lambda)=0,
\label{inst}
\ee
which can be found for arbitrary noise statistics by a
reversion of the power series,
\be
H(\lambda) = (I_L-I_R)\lambda + (1/2) (F_L+F_R)\lambda^2+\ldots=0,
\label{H}
\ee
where $I_{L,R}\equiv\la \tilde I_{L,R}\ra$ and 
$F_{L,R}\equiv\la\la\tilde I^2_{L,R}\ra\ra$.

There are two physical limits that we now
consider, based on the comparison of the maximum current difference
through the instability $\delta I = {\rm max}\{\vert I_L-I_R \vert \}$ 
(referred to as the {\it current threshold})
to the total noise power at the instability, $F=(F_L+F_R)$.  In the
{\it Gaussian} limit, discussed in the next section, the current threshold is small compared to the total noise power, $\delta I \ll
F$ so the system is effectively driven by Gaussian noise alone, with
higher current cumulants making only small corrections.  In the {\it threshold}
limit, discussed in Sec.~\ref{thresh}, the current threshold is large
compared to the total noise power, $\delta I > F$, so it will be the
tails of the distribution that drive the switch.

In the counting statistics literature, it is usually the generating
function of the stochastic process that is sought.  We would like
to comment that the instanton line (\ref{inst}) for a noiseless
nonlinear element characterizes the stochastic process in a different
way that nevertheless contains all the information about the rare
events.  Furthermore, it is directly related to a physical quantity
that is readily observed in experiments, the switching rate
(\ref{delta}), and therefore provides a more useful characterization
of the EVS.  

\section{I-V curve in the Gaussian limit}
\label{gaussian}
\begin{figure}[t]
\begin{center}
\leavevmode
\psfig{file=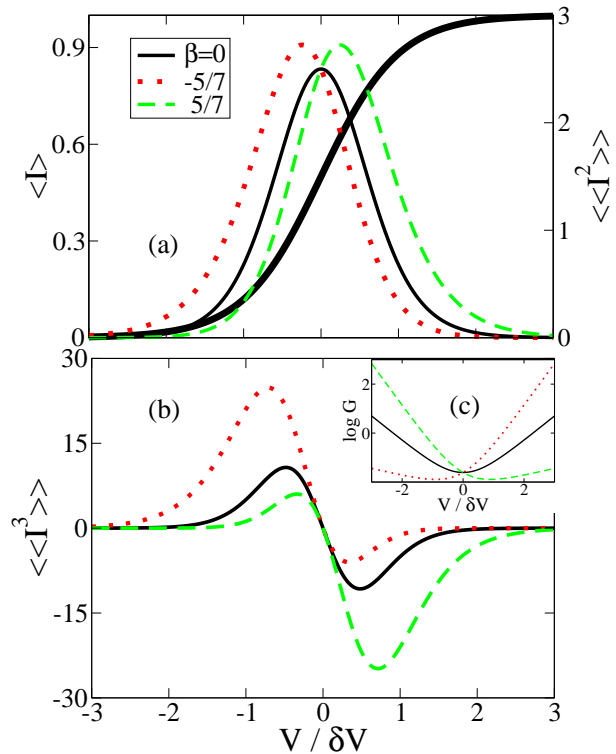,width=8cm}
\caption{(color online). The average current $\la I\ra$, the noise
$\la\la I^2\ra\ra$, and third cumulant $\la\la I^3\ra\ra$ are plotted
as a function of the scaled voltage, for different asymmetry
parameters, $\beta$.  (a) The asymmetry is invisible in the current,
but it produces a shift in the peak location of the noise.  (b) The
asymmetry is further magnified in the third cumulant, that shows
a weighted peak and dip. (c) By plotting the logarithm of the ratio
${\cal G}=(dI/dV)/\la\la I^2\ra\ra$ versus voltage, the asymmetry
parameter $\beta$ is given asymptotically by the negative difference
of the slopes, divided by the sum of the slopes.  We have taken
$I_1=0,I_2=1$, $V_0=0$, and $\Gamma_0=.1$.}
\label{step}
\end{center}
\end{figure}

We now consider the Gaussian limit, $\delta I\ll F$, and demonstrate
how to extract the noise alone from the telegraph process.
Keeping only the first two terms in Eq.~(\ref{H}), 
the instanton line is given by
\be
 \lambda_{\rm in}(Q)=-2 (I_L - I_R)/(F_L+F_R),
\label{action}
\ee 
implying that $\vert \lambda_{\rm in} \vert \ll 1$ and justifying
the series truncation.  To leading order in $\delta I/F$, the
microscopic noise is constant, $F=F_L+F_R=const$, and may be taken out
of the action integral.  For a well developed instability,
$A_{1,2}>1$, the current will make a transition from $I_1$ to $I_2$ on
a voltage scale smaller than the total instability scale
$(Q_2-Q_1)/C$ (see below).  On this smaller voltage scale, the
currents $I_{1,2}$ are approximately constant, and we may linearize
the actions $A_{1,2}$ in voltage around the point $V_0$
where they are equal, $A_1=A_2$.
This linearization gives the transition rates,
\be \Gamma_{1,2} = \Gamma_0 \exp[-(2 C/F) (I_{1,2}-I_u)(V-V_0)],
\label{rates}
\ee
where $I_{1,2}$ and $I_u$ are taken at $V=V_0$.
The rates have an activation form with (in general) different 
energy scales.  Nevertheless, the I-V curve, Eq.~(\ref{c1}), 
depends only on the ratio of the rates, and therefore has a universal form,
\be 
I(V) = \frac{I_1+I_2}{2} +
\frac{\Delta I}{2} \tanh \left[(C\, \Delta I/F)\,(V-V_0)\right].
\label{integratedcurrent}
\ee
Thus, as a function of $V$, the current has a step on a voltage scale
 $\delta V = F/(C \Delta I)$.  The conditions $A_{1,2} \sim \vert 
Q_{1,2}-Q_u\vert \delta I/F >1$ and $\Delta I >\delta I$, imply
$\delta V < (Q_2-Q_1)/C$, so the action linearization is justified.

Assuming the capacitance $C$ is known, the noise power driving 
the instability can be accurately obtained by fitting data with
Eq.~(\ref{integratedcurrent}), with $F$ as the only fitting parameter.
In contrast, the noise power and third cumulant of the telegraph
process do not have a universal form because they depend on the rates
directly.  The behavior of the cumulants may be characterized by an
asymmetry parameter $\beta=(I_1+I_2-2 I_u)/\Delta I$ that describes
the difference in the activation energy scales of the rates
(\ref{rates}).  In Fig.~2, we plot the first three cumulants,
Eq.~(\ref{c1},\ref{c2},\ref{c3}), for the rates in Eq.~(\ref{rates})
versus the normalized external voltage for different values of the
asymmetry parameter.  The asymmetry is invisible in the current,
creates a small shift in the noise peak, and is further magnified in
the third cumulant. We would like to stress that the third cumulant
may have either a peak or dip, depending on the sign of the asymmetry
parameter.  The asymmetry may be directly extracted by plotting the
logarithm of the ratio ${\cal G}=(dI/dV)/\la\la I^2\ra\ra$ versus bias
and reading off the asymptotic slope as done in Fig.~2c.  The
asymmetry parameter $\beta$ is given by the negative difference of the
slopes, divided by the sum of the slopes.

By first calibrating with an equivalent resistor to determine $F_L$
and $C$, (where the bistable system is driven by the noise of the
nonlinear system alone), the shot noise power $F_R$ of the mesoscopic sample
may be extracted.  Alternatively, one may use a nonlinear system with
known noise properties.  Note also, that the detailed shape of the
nonlinear I-V is not important, so long as the above assumptions are
met.  The accuracy of the measurement is limited by the accuracy of
the I-V curve.  The signal-to-noise ratio grows as the square
root of the number of switches, and should be large. To move to
another bias point, the nonlinear I-V curve should be shifted up.  This
can be done by attaching an additional current bias line between the
circuit elements, {\it e.g.} with a separate bias and tunnel junction.  The
external voltage and current bias allow a fully tunable bistability.
This method may be applied even for macroscopic unstable systems, such
as resonant tunneling diodes, \cite{rtw1,rtw2,rtw3} because while $C$
is large, $\delta I$ can always be made smaller by shifting the bias
to reduce the current barrier.

\section{External circuit effects}
\label{CE}
  In this section, we investigate how the external circuit can
influence the statistical properties of the bistable system.  In the
first experiment on non-Gaussian noise, feedback effects from the
external circuit played an important role.\cite{reulet,beenakker}  
If the mesoscopic system
is imperfectly voltage biased, the voltage across the mesoscopic
sample will fluctuate on a long time scale.  These slow fluctuations
alter the transport conditions, and provide additional contributions
to the individual current cumulants, named `cascade
corrections'.\cite{Nagaev1,Nagaev2,us2} We argue below that the voltage
fluctuations across the circuit may also affect the switching rates.

We begin this analysis by considering a common experimental set-up,
the current biased circuit, where the external circuit resistance is much
larger than the sample resistance. The large circuit resistor fixes
the current through the sample so current fluctuations vanish,
creating voltage fluctuations instead.  The transport dynamics is
characterized by three relevant time scales. The RC time of the system
$\tau_C$, the RC time of the external circuit $\tau_{RC}$, and the
inverse switching rate $\Gamma_S^{-1}$. 
\begin{figure}[t]
\begin{center}
\leavevmode
\psfig{file=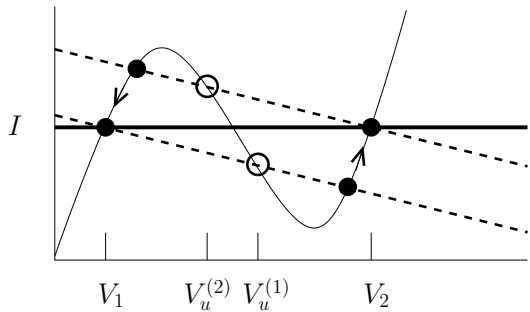,width=7cm}
\caption{System dynamics in the case of a current biased circuit.
Starting at $V_1$, the system can hop to the empty circle at
$V_u^{(1)}$ on the time scale $\tau_C$, and then fall to the black dot
on the lower dashed load line.  This creates a current below the
external current bias, so the circuit adiabatically adjusts by moving
along the arrow to $V_2$ on the time scale $\tau_{RC}$.  After a time
on the order of $\Gamma_S^{-1}$, the system can then hop over the
barrier to the empty circle at voltage $V_u^{(2)}$, and fall to the
black dot on the upper dashed load line. This state has a current
above the bias current, so the system adiabatically follows the arrow
back down to $V_1$ on the time scale $\tau_{RC}$, completing the
cycle.}
\label{currentbiased}
\end{center}
\vspace{-5mm}
\end{figure}
Any realistic measurement circuit is current biased on the external
RC-time, longer than the system relaxation time, leading to the ordering, 
\be \tau_C \ll \tau_{RC},  \Gamma_S^{-1}.
\label{real}
\ee

We first consider the experimental situation 
when the typical time spent in the stable states is
much longer than the external circuit RC-time,
\be
\tau_{RC} \ll \Gamma_S^{-1}.
\label{case1}
\ee
In this parameter range, the dynamics is sketched in Fig.~\ref{currentbiased}.
On the time scale $\tau_C$, the average voltage across the nonlinear system
changes very little, so the dynamics is effectively voltage biased.
The system transitions from $V_1$ or $V_2$
to the black dot along the slanted load line.  On the time scale
$\tau_{RC}$, the voltage across the nonlinear element adiabatically relaxes to
restore the current to its proper value.  The system then switches
again on the other slanted line after a time $\Gamma_S^{-1}$.  The main
difference with respect to the voltage biased case is that the step
will be in voltage, not current, and therefore the I-V curve will have
a plateau, not a step.  Repeating the derivation that lead to
Eq.~(\ref{integratedcurrent}), we find,
\be 
V(I) = \frac{V_1+V_2}{2} +
\frac{\Delta V}{2} \tanh \left[(C {\widetilde{\Delta V}}/F)\,
(I-I_0)\right],
\label{integratedvoltage}
\ee 
where $I_0$ is the current value where the rates are equal, and
${\widetilde {\Delta V}} = \Delta V + V^{(1)}_u - V^{(2)}_u$ is not
the same as in the prefactor because the values of the unstable
voltage are different in the shifted curves of
Fig.~\ref{currentbiased}.  
This non-universality will be small if the
mesoscopic element has a large resistance, so $V^{(1)}_u - V^{(2)}_u$
is small.  As the above analysis shows, the usual experimental
procedure of taking voltage noise data, and converting it into current
noise fails if there is an instability.

A separate circuit effect arises because the stable current state
produces its own shot noise, that the external resistor suppresses,
creating voltage noise on a time scale $\tau_{RC}$ across the
mesoscopic part of the circuit. This voltage noise adiabatically rocks
the current threshold, increasing the average transition rate.  The
relative magnitude of this effect can be estimated by comparing the
variance of the rocking potential with the voltage scale of the
transition, $\delta V$, and is small if $\tau_C \ll \tau_{RC}$, as we
have assumed.

Considering now the regime 
\be
\Gamma_S^{-1}\ll \tau_{RC}, 
\label{lang}
\ee where the telegraph switching is fast compared to the external
circuit response time, we see that the voltage across the sample has
no time to change until after the next switching event restores the
current to its original value.  In this regime, the switching is
always voltage biased.  Furthermore, we may consider the whole sample
as a fast Langevin noise source with telegraph current statistics that
drives the charge fluctuations in the external circuit.  Current
cumulants may now be computed in the usual way for stable systems with a
nonlinearity.\cite{us2,note3}

\section{Threshold Detectors and Full Counting Statistics}
\label{thresh}

\begin{figure}[t]
\begin{center}
\leavevmode
\psfig{file=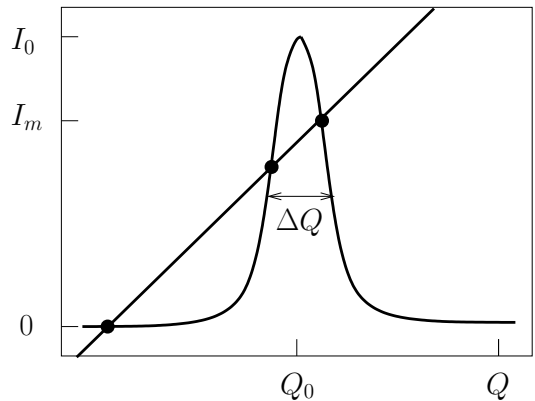,width=7cm}
\caption{Schematic of the average current flowing through the
  threshold detector and Ohmic mesoscopic conductor as a function of
  the charge.  The detector current has a peak with center $Q_0$,
  width $\Delta Q$, and maximum current $I_0$.  The mesoscopic conductor
  is defined to have currents $I=I_m$ and $I=0$ at the stable states.}
\label{IVDD}
\end{center}
\vspace{-5mm}
\end{figure}
 We now propose a measurement scheme for the EVS
of a mesoscopic conductor. The idea is to use the bistable
system as threshold detector, in the limit $\delta I > F$, where
the switch will be driven by the non-Gaussian tails of the 
current distribution.
We consider the shot noise regime, where the current cumulants 
are proportional to the average
current, $H_\alpha(Q, \lambda) = I_\alpha (Q) h_\alpha(\lambda)$,
and $h_\alpha$ generates the generalized Fano factors.  
Then, Eq.~(\ref{inst}) for the instanton line takes the following form,
\be 
-h_L(\lambda)/h_R(-\lambda) = I_R(Q)/I_L(Q) \equiv {\cal R}(Q).
\label{zeroenergy}
\ee 
In this equation, all the non-universal details of the charge
dependence of the instability appear on the rhs in the current ratio
${\cal R}$, while the statistical nature of the fluctuations appears
on the lhs.  In order to probe the probability of having a very large
(small) current in the mesoscopic system, the threshold limit we are
now concerned with implies that the extremal value of ${\cal R}$
through the instability is much smaller (larger) than 1.  This means
that in contrast to the Gaussian limit, where Eq.~(\ref{zeroenergy})
gives $\lambda_m \equiv {\rm max}\{\vert \lambda_{\rm in} \vert\} \ll
1$, the threshold limit
implies $\lambda_m > 1$.  This corresponds to large
action, or a very small switching rate, which makes the measurement
of the FCS experimentally challenging.  To overcome this difficulty a
general strategy should be based on the following ingredients:

\begin{enumerate}
\item{A separation of time scales, that allows the measurement 
of the Markovian FCS of the fast microscopic noise sources that drive
the classical circuit on a longer time scale.} 
\item{The action $A \sim \Delta Q \lambda_m $ must 
be larger than one, but not so large that the system never switches
on experimental time scales.}
\item{The instability must be such that the current ratio 
${\cal R}=I_R/I_L$ is larger (or smaller) than one in the 
bistable range, so that $\lambda_m \ge 1$.}
\item{A sufficiently large bias, so that the circuit
is both in the bistable range, and the bias across the
mesoscopic element exceeds the temperature.}
\item{A further useful (but not essential) ingredient is that
the nonlinear element is noiseless, so that the transition is
driven by the mesoscopic element alone.}
\end{enumerate}

Condition (2) is the most severe constraint.  In order to have the
action not too large, $\Delta Q$ must be comparable to the electron
charge, implying that the capacitance of the circuit is in the
mesoscopic range.  This excludes macroscopic nonlinear elements such
as tunnel diodes from measuring full counting statistics (though not
Gaussian noise, see Sec.~\ref{gaussian}).  Conditions (2) and (3)
together determine the necessary shape of the instability.  In order
to have $\Delta Q$ small, and ${\cal R}=I_R/I_L$ large, the I-Q
characteristic of the nonlinear element should have a sharp peak or
dip, the first of which is shown in Fig.~4.

To measure FCS using this peak, the switching rate should be measured
as a function of the external bias, that moves the mesoscopic load
line down the peak.  This procedure is sketched in the inset of
Fig.~5, which shows the real time switching from the stable point with
average current $I_m$, via the current peak, to the other stable point
with zero current.  The dependence of $I_m$ versus $V-V_{th}$ is shown
in Fig.~5, where $V_{th}$ is the value of the external bias at the
current maximum.  A direct measure of the current EVS may be obtained
by dividing the log-rate by the voltage jump, in order to remove
the effect of the shape of the current peak, and plotting 
\be 
S= (\log \Gamma) /C(V-V_{th}),
\label{S}
\ee 
versus $I_m/I_0$. 
\begin{figure}[t]
\begin{center}
\leavevmode
\psfig{file=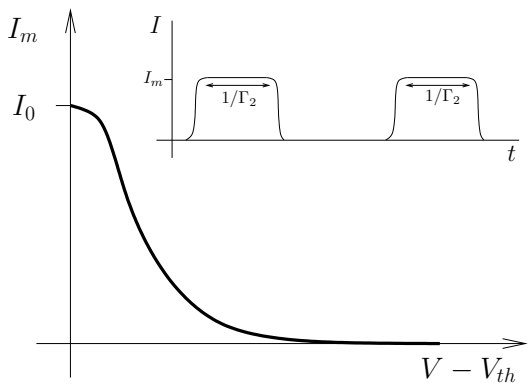,width=7cm}
\caption{Sketch of the experimental procedure to extract
the EVS of a mesoscopic conductor.
The current in the conductive state $I_m$ decays as a function
of the external bias on the scale of the peak width $\gamma$.  
Inset: The system will switch from the
conducting state to the insulating state on a time scale $1/\Gamma_2$}
\label{exp}
\end{center}
\vspace{-5mm}
\end{figure}

We now consider the switching rate
for Poissonian, Gaussian, and Binomial noise sources
while assuming a noiseless nonlinear element ($h_L = \lambda$)
so that the switching dynamics is governed solely by the system noise.
The characterization of EVS can be
extracted from the divergence of the action, which corresponds to
large values of $\lambda_m$.  We compare the most important
processes: Gaussian, $h_R(\lambda) = \lambda + \lambda^2/2$ (Fano
factor is 1); Poissonian, $h_R(\lambda) = \exp \lambda -1$; and
Binomial, $h_R(\lambda) =T^{-1}\log[1+T(\exp \lambda -1)]$.  In the
asymptotic limit ($\vert \lambda \vert\rightarrow \infty$), 
Eq.~(\ref{zeroenergy}) may be solved to obtain: 
\be
\lambda_{\rm in} = \begin{cases} - 1/{\cal R}, & {\rm Gaussian} \cr 
\log {\cal R}, & {\rm Poissonian} \cr 
(T \log T)/({\cal R}-T), & {\rm Binomial}
\end{cases}
\label{lam}
\ee
which replaces Eq.~(\ref{action}).

\begin{figure}[t]
\begin{center}
\leavevmode
\psfig{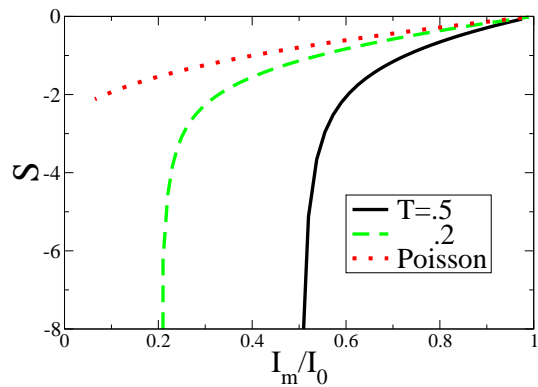}
\caption{(color online). The normalized log-rate $S$ is
plotted versus $I_m/I_0$ for a quantum point contact with
transparencies $T=.5$ and $T=.2$ measured by a noiseless detector with
current peak of Lorentzian shape.  $S$ has a power-law divergence as
$I_m/I_0$ approaches $T$, which is a manifestation of the `Pauli
stabilization' effect (see text).  Comparison is also shown with the Poissonian
limit, whose divergence at the origin is cut off by the finite
conductance of the QPC.}
\label{rate}
\end{center}
\end{figure}
It is important that all three processes have very different
asymptotic behaviors that makes it relatively easy to distinguish 
them in experiments.   However, the most surprising fact is that the
Binomial process, characteristic of a quantum point contact with
transparency $T$,
 has a sharp power-law singularity at ${\cal R}=T$.  It
persists even at small $T$ (the tunneling limit) which is usually
considered to give a Poissonian process.
This behavior (discussed previously by Tobiska and Nazarov in 
Ref.~\onlinecite{nazarovJJ}) has the following physical
interpretation:  The total
charge that passes the conductor is the sum of independent electron
attempts, with success probability $T$, and failure probability $1-T$.
The Pauli principle allows only one electron at a time to make an
attempt.  Therefore, the current distribution has a sharp cut-off at
the maximum allowed current, when all attempts are successful.  This
maximum current is given by $I_{max} = \la I\ra /T$.  If the current
threshold ratio $\cal R$ is lowered below $T$, the mesoscopic
conductor has no chance to have a large enough fluctuation to overcome
the barrier, and the system never switches.  We propose the name
``Pauli stabilization'' to describe this impotency.  To
further illustrate the effect, we plot $S$
Eq.~(\ref{S}) for a Lorentzian peak in Fig.~6, using a quantum point
contact with different transparencies.  Even if the transparency is
fairly small, where Poissonian statistics is naively expected, there
is still a power-law divergence in $S$.  This divergence may be estimated by
expanding ${\cal R }\approx {\cal R}_0 + \alpha Q^2$ near the peak,
to obtain
\be
S \sim -(T\log T)/\sqrt{\alpha ( {\cal R}_0 -T)}.
\label{act}
\ee

An interesting situation occurs when the bistable system is driven
by a microscopic noise that is itself a random telegraph process.
For instance, a charge trap near the right mesoscopic conductor may switch
the current between $I_a$ and $I_b$, with rates $\Gamma_{a,b}$.
The generating function of this random process is\cite{us0}
\begin{eqnarray} 
H_{R} &=& \frac{1}{2}(I_a +I_b)\lambda -\frac{1}{2} 
(\Gamma_a +\Gamma_b)\nonumber \\
&+&\sqrt{[(I_b-I_a)\lambda - \Gamma_b + \Gamma_a]^2/4 +\Gamma_a
\Gamma_b}.
\label{rtn}
\end{eqnarray}
The instanton line (\ref{inst}) has an exact solution
for a noiseless nonlinear element,
\be
\lambda_{\rm in}(Q) = \frac{\Gamma_a}{ I_L - I_a } 
+ \frac{\Gamma_b}{I_L - I_b} .
\label{RTNin}
\ee 
An important check is $\lambda_{\rm in} = 0$ when
$I_L=I_R =(\Gamma_b I_a + \Gamma_a I_b)/(\Gamma_a+\Gamma_b)$.  On the
other hand, the instanton solution diverges when $I_L$ approaches
$I_a$ or $I_b$, where the distribution has a cut-off,\cite{us0} and
thus also displays a stabilization effect.  As $I_L$ approaches $I_{a}$
or $I_b$, the currents may be approximated as $I_{b} - I_L \approx
(I_{b}^0 - I_L^0) + \epsilon\, Q^2$, so the action itself has a 
power-law divergence as 
\be S \sim
-\Gamma_{a,b} /\sqrt{\epsilon\, (I_b^0 - I_L^0)}.
\label{div}
\ee 

It is important to note that because the telegraph process has a much
larger noise power than shot noise, this stabilization effect should be able
to be observed even with macroscopic nonlinearities, such as tunnel
diodes.  To see why this is so, we estimate the action away from the
divergence as $S \sim \Gamma_{a,b}\, \tau_C$.  Our time scale separation
demands that $\Gamma_{a,b} >\tau_C$, so $S>1$.  Other than this
requirement, $\Gamma_{a,b}$ is an independent parameter.  Therefore,
the action can be made of order one even with a large capacitance, so
the action divergence from the EVS stabilization behavior in
Eqs.~(\ref{RTNin},\ref{div}) should be able to be seen on experimental time
scales.

\section{Double quantum dot as a threshold detector}
\label{DD}
While the discussion has thus far been rather general, we now
concentrate on a specific implementation of the mesoscopic threshold
detector, a quantum double dot (DD).\cite{DDreview} We consider the
resonant tunneling regime, where each dot has a Breit-Wigner resonance
of Lorentzian shape.\cite{stone,markus} 
The transmission as a function of energy is 
\be
T(E) = \frac{4 \gamma^2 t^2}
{\vert (E - E_0 + i \gamma)(E + E_0 + i \gamma)\vert^2},
\label{t}
\ee where $\gamma$ is the total decay width of the symmetric resonant
levels, $t$ is the tunnel coupling of the middle barrier, $E_0^2 = t^2
+ (\Delta E)^2/4$ is the hybridized energy of the levels, and $\Delta
E$ is the energy difference between the levels that can be adjusted
with gate voltages.
The total current through the
double dot also has a Lorentzian shape as a function of $\Delta E$, 
\be I = \frac{2 \pi \gamma t^2}{\gamma^2 + E_0^2},
\label{bw}
\ee and is maximal at $I_{\rm max} = \pi \gamma$ for $\gamma=t$, and
$\Delta E =0$.  Despite the fact that the transparencies of the
barriers are low, the total current is large in the resonant tunneling
limit.  In Fig.~4, $\Delta E \propto Q$, where the coefficient is
relative capacitance of the two levels to the cavity, and $\Delta Q
\propto \gamma$.  The Fano factor $f$ at $\Delta E=0$ as a function of
the dimensionless ratio $r$ is 
\be f = \frac{2 r^4
-r^2+1}{2 (1 + r^2)^2}, \quad r = \gamma/t,
\label{fanoscale}
\ee and has a minimum at $r=\sqrt{3/5}$, where the Fano factor is
$f_{\min} = 7/32 \approx .219$.  At this point, the noise is
suppressed, but not zero.  It is well known that a single Breit-Wigner
resonance has a Fano factor of $1/2$, and here with a double-dot,
it is suppressed below $1/4$.  These results naturally lead to 
the idea that a series of quantum dots, or a ballistic narrow-band conductor, 
may be used as an ideal noise detector.

\begin{figure}[t]
\begin{center}
\leavevmode
\psfig{file=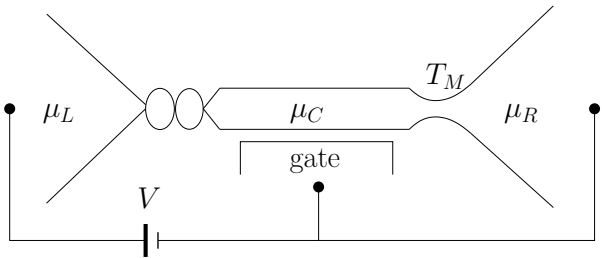,width=8cm}
\caption{A double quantum dot in series with a
mesoscopic quantum point contact with transmission $T_M$.  A metallic
side gate provides a tunable capacitance.}
\label{DDcircuit}
\end{center}
\vspace{-5mm}
\end{figure}

This double dot is fabricated together with a quantum point contact
(QPC) connected through a mesoscopic cavity as sketched in
Fig.~\ref{DDcircuit}.  The physics of the switch is as follows.
Current is flowing through the DD with the right level slightly above
the left, and flowing out the QPC.  The QPC has a rare event, where
many subsequent electrons succeed in exiting the right contact.  This
depresses the charge in the cavity below the average, lowering the
potential on the cavity.  The potential capacitively couples
asymmetrically to the two quantum dots which aligns the levels on the
DD, producing more current flowing into the cavity.  The QPC continues
in its rare event, further lowering the potential in the cavity,
finally misaligning the levels to the unstable point, and cutting off
transport.

We now make some estimates of the energy scales and parameter
ranges for the circuit to function as we wish.  There should only be a
few resonant levels in the transport window, so the typical energy
spacing between the adjacent peaks in the I-V curve will be the mean
level spacing of the quantum dots, $\Delta_{D}$.  The width of the
current peak is $\gamma < \Delta_D$,  implying the peaks are
separated.  The current at the top of the peak is given by the peak
conductance times the width of the barrier, $I_0 \le \gamma$,
where equality is reached in the perfect resonant tunneling limit.  

The first condition is on the conductance of the mesoscopic
sample $T_M$ (we set the conductance quantum equal to 1), 
so the load line crosses one peak only, as shown in Fig.~4,   
\be
I_0/\gamma > T_M > I_0/\Delta_{D}.
\label{cond}
\ee 
The left inequality is most strict for an open QPC, which requires
perfect resonant tunneling. The right inequality is 
not very restrictive, since $I_0/\Delta_{D} = (I_0/\gamma)(\gamma/\Delta_D)<1$,
which allows the pinch-off limit.

The next condition is the time scale separation between the RC-time of
the cavity, and the tunneling time through the DD.  In the
case, $C > \Delta_C^{-1}$, where $C$ is the geometrical
capacitance, and $\Delta_C$ is the mean level spacing in the
connecting cavity at the Fermi energy, $C_\mu^{-1}=(C^{-1}+\Delta_C)
\approx \Delta_C$.  The time scale separation condition 
$\gamma > \Delta_C T_M$, simply means that the cavity is larger
than the quantum dots.  Additionally, the action must not be too large, so
for the FCS measurement (where $\lambda_m > 1$), the charge
difference, $\Delta Q$, must not be too large.  The charge width is 
given by the density of states of the cavity, times the peak's voltage width, 
$\Delta Q \sim \gamma/\Delta_{C}$.  Together, these two conditions
constrain the size of the cavity,
\be
\gamma/T_M > \Delta_C > \gamma/\Delta Q,
\label{dc}
\ee
which gives a rather small parameter range.

Finally, to go out of equilibrium, the system must be cooled to
mesoscopic temperatures, $T < I_0/T_M$, much smaller than the mean
level spacing of the small quantum dot.  However, it would be also
interesting to observe EVS even in equilibrium.  While the noise power
in equilibrium is simply a consequence of the fluctuation-dissipation
theorem, the EVS is nontrivial.

We now compare our idea with other proposals for
measuring noise and EVS of mesoscopic conductors.
In the proposals discussed below, the measurement device is in
a separate circuit weakly coupled to the mesoscopic conductor
that acts as an external noise source.
Aguado and Kouwenhoven proposed a double quantum dot as
a detector of high frequency noise.\cite{noiseDD}  The quantum noise causes
inelastic transitions between the states of quantum dot, and therefore
the double dot current is proportional to the noise power at that
frequency.  
Tobiska and Nazarov proposed using Josephson junctions
as a measurement device.\cite{nazarovJJ}  A rare current fluctuation causes the
quantum phase to jump over the top of its effective potential,
creating a transition from the superconducting phase
to the normal conducting phase.  The switching rate gives
information about the probability of the rare current fluctuation.
Macroscopic quantum tunneling is suppressed by having an array
of Josephson junctions to make the potential barrier width large.
Pekola also proposed using Josephson junctions as 
a measurement device.\cite{PekolaJJ}  In this proposal, the noise first 
creates a transition from the ground state to the first excited state,
where macroscopic quantum tunneling causes escape.

The noise experiment implementing this last proposal,
Ref.~[\onlinecite{pekolaexp}], provided additional clarification of
the difficulties involved in measuring FCS.  Similarly to the
double-dot detector, \cite{noiseDD} the Josephson junction measured
noise at the plasma frequency of the junction, while the low frequency
noise leaked through the bias line.  This also explained why the
expected exponential dependence of the rate on the current threshold
was not found.

Our proposal is based on essentially different physics: The threshold
detector is supposed to work in a regime where detection is a
non-perturbative process due to strong coupling to the measured system.
Although a realization of this threshold detector is an experimental
challenge, there are several advantages of our proposal.  The
separation of time scales requirement $\tau_C \gg \tau_0$ is necessary
to measure low frequency noise:  The finite response time allows many
electrons to enter and leave the cavity, so the Markovian limit is
reached.  This limit also implies that quantum effects are not
relevant.  Detector feedback, usually a liability, is completely
accounted for.  In fact, detector feedback is an essential ingredient
for our proposal.

\section{Conclusions}
\label{conc}

We have proposed the use of circuit instabilities as two-terminal
detectors of low-frequency noise.  The considered circuit consists of a
mesoscopic conductor with parallel capacitor, in series with a
nonlinear element.  The nonlinear device contains a region of negative
differential resistance, which allows bistability.  There are two
regimes of interest, from the point of view of shot noise measurement.

The first is the Gaussian regime, where the noise power is much larger
than the current threshold.  In this limit, the noise power is effectively 
constant, and the higher cumulants may be neglected.  The noise 
drives a transition between two current values, the telegraph 
process.  This process
produces a step in the I-V curve of universal form, with only one
variable parameter, from which the noise power may be extracted.  The
second cumulant has a peak at the current step, while the third
cumulant has a peak and a dip, the relative weight depending on an
asymmetry parameter of the switching rates.  We further considered
external circuit effects.  In the current biased case, the dominant
effect is that the I-V curve has a plateau, not a step, because it is
the voltage that switches, not the current.  The measurement of
Gaussian noise may be carried out with macroscopic conductors
containing nonlinearities, such as resonant tunneling wells.

The second regime is the threshold regime, where the noise power is
smaller than the current threshold.  In this limit, the switch comes
from the tails of the distribution, and is a direct signature of the
extreme value statistics of the mesoscopic conductor.  The most
interesting effect occurs for charge distributions that have a
cut-off.  This cut-off manifests itself in a divergence of the
switching rate, that stabilizes the state.  We considered both Pauli
stabilization from a quantum point contact, as well as stabilization
from a microscopic random telegraph process.  While the measurement of
full counting statistics requires a mesoscopic instability because of
the long time scales involved, the stabilization effect from the
random telegraph process should be visible in macroscopic nonlinear
elements.  We proposed a quantum double dot operating in the resonant
tunneling regime as an implementation of the threshold detector of rare
shot noise fluctuations.  Constrains on the conductance of the
measured conductor, and the capacitance of the central dot were
discussed.

\acknowledgments
We thank M. B\"uttiker, C. Sch\"onenberger, and M. Reznikov for discussions.
This work was supported by MaNEP and the Swiss National Science Foundation.


\begin{thebibliography}{02}



\bibitem{BB}
Ya. M. Blanter and M. B\"uttiker,
Physics Reports {\bf 336}, 1 (2000).

\bibitem{BS}
C.W.J. Beenakker and C. Sch\"onenberger,
Physics Today, {\bf 56}, 37 (2003).

\bibitem{reulet}
B. Reulet, J. Senzier, and D. E. Prober,
Phys. Rev. Lett. {\bf 91}, 196601 (2003).

\bibitem{R}
M. Reznikov {\it et al.}, unpublished experiment.

\bibitem{FCS1}
L. S. Levitov and G. B. Lesovik,
Pis'ma Zh. Eksp. Teor. Fiz.
{\bf 58}, 225 (1993).

\bibitem{FCS2}
L. S. Levitov, H. Lee and G. B. Lesovik,
J. Math. Phys. {\bf 37}, 4845 (1996).

\bibitem{FCS3} 
{\it Quantum Noise in Mesoscopic Systems}, edited by Yu. V. Nazarov
NATO Science Series II Vol. 97 (Kluwer, Dordrecht, 2003).



\bibitem{nazarovJJ}
J. Tobiska and Yu. V. Nazarov,
Phys. Rev. Lett. {\bf 93}, 106801 (2004).

\bibitem{us0}
A. N. Jordan and E. V. Sukhorukov,
Phys. Rev. Lett.  {\bf 93}, 260604 (2004).

\bibitem{SB}
E. V. Sukhorukov and O. M. Bulashenko, 
Phys. Rev. Lett. {\bf 94}, 116803 (2005). 



\bibitem{PekolaJJ}
J. P. Pekola,
Phys. Rev. Lett. {\bf 93}, 206601 (2004).

\bibitem{onchip}
Y. Nakamura, Yu. A. Pashkin, and J. S. Tsai, 
Nature (London), {\bf 398}, 786 (1999).

\bibitem{rmp}
Y. Makhlin, G. Sch\"on, and A. Shnirman,
Rev. Mod. Phys. {\bf 73}, 357 (2001).


\bibitem{noiseDD}
R. Aguado and L. P. Kouwenhoven,
Phys. Rev. Lett.  {\bf 84}, 1986 (2000).

\bibitem{schoelkopf}
R. J. Schoelkopf, A. A. Clerk, S. M. Girvin, K. W. Lehnert, and
M. H. Devoret, in Ref.\onlinecite{FCS3}.

\bibitem{pekolaexp} 
J.P. Pekola, T.E. Nieminen, M. Meschke,
J.M. Kivioja, A.O. Niskanen, J.J. Vartiainen, cond-mat/0502446.


\bibitem{machlup}
S. Machlup, J. Appl. Phys. {\bf 25}, 341 (1954).




\bibitem{note1}
If the rates are not that small, it is important to
first subtract off the noise background in Eqs.~(\ref{c2},\ref{c3})
before applying the self-consistent check.  This is straightforward to
do, because the crossover between the stable states in higher
order cumulants is also universal.

\bibitem{shuttle}
C. Flindt, T. Novotny, and A.-P. Jauho,
Europhys. Lett., {\bf 69}, 475 (2005).


\bibitem{us1}
S. Pilgram, A. N. Jordan, E. V. Sukhorukov, and M. B\"uttiker,
Phys. Rev. Lett. {\bf 90}, 206801 (2003).

\bibitem{us2}
A. N. Jordan, E. V. Sukhorukov, and S. Pilgram,
J. Math. Phys. {\bf 45}, 4386 (2004).

\bibitem{spi1}
S. Pilgram, Phys. Rev. B {\bf 69}, 115315 (2004).

\bibitem{spi2}
K. E. Nagaev, S. Pilgram, and M. B\"uttiker, 
Phys. Rev. Lett. {\bf 92}, 176804 (2004).
\bibitem{spi3}
S. Pilgram, K. E. Nagaev, and M. B\"uttiker, Phys. Rev. B {\bf 70}, 045304
 (2004).
\bibitem{spi4}
 S. Pilgram, P. Samuelsson, Phys. Rev. Lett. {\bf 94}, 086806 (2005).


\bibitem{rtw1}
Ya. M. Blanter and M. B\"uttiker,
Phys. Rev. B {\bf 59}, 10217 (1999).

\bibitem{rtw2}
O. A. Tretiakov, T. Gramespacher, and K. A. Matveev,
Phys. Rev. B {\bf 67}, 073303 (2003).

\bibitem{rtw3}
V. V. Kuznetsov, E. E. Mendez,  J. D. Bruno, and J. T. Pham,
Phys. Rev. B {\bf 58}, 10159(R) (1998).

\bibitem{beenakker}
C.W.J. Beenakker, M. Kindermann, and Yu. V. Nazarov, 
Phys. Rev. Lett. {\bf 90}, 176802 (2003).

\bibitem{Nagaev1}       
K. E. Nagaev,
Phys. Rev. B {\bf 66}, 075334 (2002).

\bibitem{Nagaev2}       
K. E. Nagaev, P. Samuelsson, and S. Pilgram,
Phys. Rev. B {\bf 66}, 195318 (2002).

\bibitem{note3} There is a further possibility that occurs when the
standard deviation of the voltage fluctuations exceed the range of the
nonlinear voltage step.  Here the linearization procedure breaks down
again, which demands a non-perturbative treatment of the entire
circuit.  This possibility only occurs in a small parameter range, and
is beyond the scope of this paper.


\bibitem{DDreview}
W.G. van der Wiel, S. De Franceschi, J.M. Elzerman, T. Fujisawa,
S. Tarucha, and L.P. Kouwenhoven,
Rev. Mod. Phys. {\bf 75}, 1 (2003).


\bibitem{stone}
A.D. Stone and P.A. Lee,  Phys. Rev. Lett. {\bf 54}, 1196 (1985).

\bibitem{markus}
M. B\"uttiker, IBM J. Res. Develop. {\bf 32}, 63 (1988).


\end{thebibliography}
\end{document}